\documentclass[%
 aip,
 amsmath,amssymb,
 reprint,%
]{revtex4-1}

\usepackage{graphicx}
\usepackage{dcolumn}
\usepackage{bm}
\usepackage{accents}

\usepackage{hyperref}

\usepackage[utf8]{inputenc}
\usepackage[T1]{fontenc}
\usepackage{mathptmx}
\usepackage{etoolbox}
\usepackage{subfig,wrapfig}

\makeatletter
\def\@email#1#2{%
 \endgroup
 \patchcmd{\titleblock@produce}
  {\frontmatter@RRAPformat}
  {\frontmatter@RRAPformat{\produce@RRAP{*#1\href{mailto:#2}{#2}}}\frontmatter@RRAPformat}
  {}{}
}%
\makeatother
\begin{document}

\preprint{AIP/123-QED}

\title[An E \& B Gyrokinetic Simulation Model for Kinetic Alfv\'en Waves in Tokamak Plasmas]{An E \& B Gyrokinetic Simulation Model for Kinetic Alfv\'en Waves in Tokamak Plasmas}


\author{M.H. Rosen}
 \altaffiliation[Present affiliation: ]{Princeton Plasma Physics Laboratory, Princeton, NJ 08543-0451, United States}
 
\author{Z.X. Lu$^*$}%
\email{zhixin.lu@ipp.mpg.de}


\author{M. Hoelzl}
\affiliation{ 
Max Planck Institute for Plasma Physics, Garching b.M., Germany
}%

\date{\today}

\begin{abstract}

The gyrokinetic particle simulation serves as a powerful tool for the studies of transport, nonlinear phenomenon, and energetic particle physics in tokamak plasmas. While most gyrokinetic simulations make use of the scalar and vector potentials, a new model (GK-E\&B) has been developed by using the E and B field in a general and comprehensive form and has been implemented in simulating kinetic Alfv\'en waves in uniform plasma [Chen et al, Science China Phys. Mechanics \& Astronomy 64 (2021)]. 
In our work, the Chen et al. GK-E\&B model has been expressed in general tokamak geometry explicitly using the local orthogonal coordinates and general tokamak coordinates. Its reduction to the uniform plasma is verified and the numerical results show good agreement with the original work by Chen et al. The theoretical derivation of the dispersion relation and numerical results in the local model in screw pinch geometry are also in excellent agreement. Numerical results show excellent performance in a realistic parameter regime of burning plasmas in terms of high values of $\beta /( M_e k_\perp^2 \rho_i^2)$, which is a challenge for traditional methods due to the ``cancellation'' problem.
As one application, the GK-E\&B model is implemented with kinetic electrons simulated using the Particle-in-Fourier method in the local single flux surface limit. With the matched ITPA-TAE parameters adopted, numerical results show the capability of the GK-E\&B in treating the parallel electron Landau damping for realistic tokamak plasma parameters. As another application, the global GK-E\&B model is implemented with the dominant electron contribution to the $E_\|$ equation in the cold electron limit. Its capability in simulating the finite $E_\|$ due to the finite electron mass is demonstrated. 



\end{abstract}

\maketitle

\section{\label{sec:level1}Introduction\protect}


The gyrokinetic particle simulation has been a powerful tool for studying tokamak plasmas in the past several decades \cite{lee1983gyrokinetic,lin1998turbulent,lanti2019orb5,hatzky2019reduction}. In simulating the electromagnetic waves and instabilities, the treatment of kinetic electrons is a challenge due to the small electron-to-ion mass ratio. Various schemes have been developed to solve this challenge, including, but not limited to, the iterative scheme for the Amp\'ere's law \cite{chen2003deltaf}, the conservative scheme \cite{bao2018conservative}, the implicit scheme \cite{Lu2021,sturdevant2021verification} and the mixed variable/pullback scheme \cite{mishchenko2014pullback,hatzky2019reduction}. The numerical challenge at high value of $\beta_e m_i/(m_ek_\perp^2\rho_{ti})$ was originally termed the ``cancellation problem'' and makes gyrokinetic simulations computationally expensive, especially for high parameter tokamak plasmas such as ITER \cite{hayward2021global}, where $\beta_e$ is the electron beta, $m_{i,e}$ is the ion or electron mass, $k_\perp$ is the perpendicular wave vector, $\rho_{ti}$ is the thermal ion Larmor radius. 

While most gyrokinetic simulations adopt the scalar and vector potentials as the field variables, a gyrokinetic simulation model has been proposed recently by making use of the electric and magnetic fields directly \cite{Chen2021}. By separating the parallel and perpendicular (to $\mathbf{B}$) dynamics, the fast electron response to the parallel electric field in the parallel direction is treated explicitly in the parallel Amp\'ere's law. This so called E and B scheme is applicable to not only the low frequency ($\omega\ll\omega_{ci}$) problem, but also the higher frequency ($\omega\ll\omega_{ce}$) problem \cite{Chen2019}. The application of this scheme to the kinetic Alfv\'en wave simulation demonstrates its capability in high $\beta_e m_i/(m_ek_\perp^2\rho_{ti})$ regime but only uniform plasma has been considered numerically. In this work, this E and B scheme is studied theoretically and numerically in tokamak geometry and screw pinch configuration. Flux coordinates and the field centered coordinates (local orthogonal coordinates) are adopted for the derivation. Two models with specific simplifications in torus geometry have been defined, namely, the local single flux surface model with kinetic electrons (local GK-E\&B model) and the global model with the dominant electron response in the zero pressure limit (global GK-E\&B fluid model). The code is developed and Alfv\'en waves are simulated, for demonstrating its performance in the simulation of tokamak plasmas. 

This article is organized as follows. In Section \ref{sec:thoery}, the base equations of the E and B scheme are given and its representation in general tokamak geometry is presented, as well as the reduction in screw pinch and ad hoc tokamak geometry. In Section \ref{sec:numerics}, the numerical scheme and the benchmark results are shown. In Section \ref{sec:results}, the simulation results in ad hoc tokamak geometry and screw pinch configuration are detailed. In Section \ref{sec:conclusion}, we give the conclusion and outlook.

\section{Theoretical Foundation}
\label{sec:thoery}
\subsection{Physics equations}
The theoretical framework for this study is laid out in Reference \onlinecite{Chen2021}. The authors present a set of five equations to describe electromagnetic fluctuations with frequencies much lower than the ion cyclotron frequency in a plasma confined by a magnetic field $\bf B (\bf r, \rm t)$ with electric field $\bf E( r, \rm t)$, fluid velocity $\bf U(\bf r, \rm t)$, current $\bf J(\bf r, \rm t)$, and pressure $\bf P(\bf r, \rm t)$,
\begin{align}
    &\partial_t\left(\rho_m \mathbf{U}_{i \perp}\right)=\frac{1}{c} \mathbf{J} \times \mathbf{B}-[\nabla \cdot \mathbf{P}]_{\perp}\;\;, \label{eq:ChenMassflow}\\
    &\mathbf{J}=\frac{c}{4 \pi} \nabla \times \mathbf{B}\;\;, \label{eq:ChenCurrent}\\
    &\mathbf{E}_{\perp}=-\frac{1}{c} \mathbf{U}_{i \perp} \times \mathbf{B}+\frac{1}{\rho_{m}} \sum_{j \neq e} \frac{m_{j}}{q_{j}}\left[\nabla \cdot \mathbf{P}_{j}\right]_{\perp}\;\;, \label{eq:ChenEperp}\\
    &c^{2}\left[\nabla_{\perp}^{2} E_{\|}-\mathbf{b} \cdot \nabla\left(\nabla \cdot \mathbf{E}_{\perp}\right)\right]=4 \pi \partial_{t} J_{\|}\;\;, \label{eq:ChenEparallel}\\
    &\partial_{t} \mathbf{B}=-c \nabla \times \mathbf{E}\;\;, \label{eq:ChenFaraday}
\end{align}
where $\rho_m$ is the mass density, subscript $i$ indicates the ion species which can contain multiple species, $c$ is the speed of light, and $q$ is charge. The parallel direction is defined as $\hat{\textbf{b}} = \textbf{B} / B$ and the components perpendicular to $\hat{\bf b}$ are denoted as $\perp$. Some of the essential assumptions made in this model are that quasi-neutrality is valid, mass of electrons is negligible compared to the mass of ions $m_e \ll m_i$, perpendicular momentum is only carried by the ions, and the mode frequency is much lower than the ion cyclotron frequency. Using this model, at each time step we must be given $\mathbf{B}$ and $\mathbf{U}$, then we can calculate the other field variables and push $\mathbf{B}$ and $\mathbf{U}$ in time. The kinetic closures to these equations comes from the terms $\rho_{m,j}=m_j n_j$, $\bf P$ and $\partial_t J_\|$,
\begin{align}
    &n_j = \langle f_j\rangle_v\;\;, \\
    &\mathbf{P}_j = \mathbf{P}_{pol,j} + \mathbf{P}_{g,j}\;\;,\label{eq:ChenPj}\\
    &\mathbf{P}_{\text {pol }, j} \simeq-(3 / 4) \mathbf{I} \nabla \cdot\left[\left(\left(N_{j} q_{j}\right) \rho_{t j}^{2} / 2\right) \mathbf{E}_{\perp 0}\right]\;\;, \label{eq:ChenPpol}\\
    &\partial_{t} J_{j \|}=q_{j}\left\langle J_{0}\left[F_{g} \partial\left(v_{\|} \dot{v}_{\|}\right) / \partial v_{\|}-\dot{\mathbf{X}} \cdot \nabla\left(v_{\|} F_{g}\right)\right]\right\rangle_{j, v} \;\;,\label{eq:ChenJparallel}
\end{align}
where $\mathbf{E}_\perp$ is solved order by order, namely, $\mathbf{E}_\perp=\mathbf{E}_{\perp 0}+\mathbf{E}_{\perp 1}+\ldots$, $\mathbf{P}_{pol,j}$ is the polarization pressure due to particle species $j$, $\mathbf{P}_{g,j}$ is the pressure from the gyrocenter, $\rho_t$ is the thermal gyroradius, $J_0 = J_0 (k_\perp \rho)$ is the Bessel function accounting for finite-Larmor-radius effects, $F_g$ is the gyrocenter distribution function, and $\langle \hdots \rangle_{j,v}$ is the velocity-space integral for species $j$. The equations of motion of the gyrocenter are as follows.

The motion of the guiding centers can be described using the symplectic or Hamiltonian formulation \cite{littlejohn1983variational,hatzky2019reduction}. In this work, in order to minimize the technical complexity, the equations of motion are adapted by keeping the dominant terms as we adopted previously \cite{Lu2021},
\begin{eqnarray}
	\label{eq:dRdt}
	 \dot{\mathbf{X}} &=& {v}_{\parallel}+{\bm v}_d+\delta{\bm v}\;\;,\\
	\label{eq:dvpardt}
	\dot{v_\parallel} &=& \dot{v}_{\parallel0} + \delta\dot{v}_{\parallel}\;\;, \\
	\label{eq:vd}
	{\bm  {v}}_d &=& \frac{m_s B_0}{ {e}_sB^2} {\rho}_N\left( {v}^2_\parallel+ {\mu} B\right)\hat{\bm b}\times\nabla B \;\;, \\
	\delta{\bm v} &=& B_0 \rho_N \left( \delta E_\perp \times \frac{\hat{\bm b}}{B} +\frac{v_\parallel}{B}\delta{\bm B}\right)\label{eq:deltavbf_torus}\\
	\dot{ {v}}_{\parallel0} &=& - {\mu}\partial_{\parallel}B\;\;,\\
	\label{eq:dvpardt_torus}
	\delta\dot{ {v}}_{\parallel} &=& \frac{\bar{e}_s}{\bar{m_s}}\delta E_\| \;\;,
\end{eqnarray}
where $\bar{m_s}$ is the mass of species $s$ normalized to the mass of the electron, $\overline{e_s}$ is the charge of species $s$ normalized to the elementary charge, the magnetic moment $\mu = v_\perp^2 / (2 B)$, and $\rho_N$ is the normalized thermal Larmor radius. More rigorous models with higher order corrections in $\rho/L_{Eq}$ (the ratio between the Larmor radius and the characteristic length of equilibrium magnetic field) can be found in other work \cite{littlejohn1983variational,white2013theory}.

\subsection{Representation in tokamak geometry}
\label{subsec:rep_torus}
While the E and B scheme has been developed in a general and comprehensive form and tested numerically in uniform plasma \cite{Chen2019,Chen2021}, the corresponding expression and implementation in tokamak geometry have not been reported. In this section, we focus on deriving the general form of the E and B scheme in tokamak geometry by choosing the magnetic flux coordinate system and specific decomposition of the field variables in the perpendicular and parallel direction. The general representation in tokamak geometry can be readily reduced to that in the screw pinch configuration, as shown in Section \ref{subsec:screw_pinch_results} and to that in concentric circular tokamak geometry, as shown in Section \ref{subsec:EBfluid_circular_torus_results}.

\subsubsection{Coordinate system and field representation}\label{sec:coordinateSystem}
Two sets of coordinate system, namely the magnetic flux coordinates and the local orthogonal coordinates, are chosen for the E and B scheme in torus geometry.
The toroidal coordinates $(r,\phi,\theta)$ are chosen for the operators such as the divergence and the gradient, where $r$, $\phi$, and $\theta$ are the radial-like coordinate, poloidal angle, and toroidal angle respectively. Generally, the radial-like coordinate can be defined as a function of the magnetic flux function, e.g., $r(\psi) = \sqrt{(\psi - \psi_{axis})/(\psi_{edge} - \psi_{axis})}$, where $\psi$ is poloidal magnetic flux (see Refs. [\!\!\!\!\citenum{Lu2012,white2013theory}] and references therein). Using this convention, we define $\mathbf{B_0} = \nabla \psi \times \nabla \phi + F \nabla \phi$, where $F$ is the poloidal current function. The Jacobian $J = \left(\nabla r \times \nabla \phi \cdot \nabla \theta\right)^{-1}$. The operators in Eqs.\ref{eq:ChenMassflow}--\ref{eq:ChenFaraday} can be readily expressed in the toroidal coordinates, e.g., $\nabla f=\nabla\alpha \partial_\alpha f$, $\nabla\cdot\mathbf{A}=(1/J)\partial_\alpha(\nabla\alpha\cdot\mathbf{A})$, where $\mathbf{A}=A_\alpha\nabla \alpha$, $\alpha\in(r,\phi,\theta)$ and Einstein summation convention is adopted.

Since the field variables are decomposed in the perpendicular and parallel directions, the second coordinate system, the local orthogonal coordinates $\hat{\mathbf{ r}},\hat{\mathbf{b}},\hat{\mathbf{\phi}}$, is adopted, where $\hat{\mathbf{r}}$ is the unit vector in the radial direction, the parallel unit vector is defined as $\hat{\mathbf{{b}}} = {\bf{B}}/{B} = \left[\nabla \psi \times \nabla \phi + F \nabla \phi\right]/B$ and $\mathbf{e}_\chi\equiv\mathbf{e}_r\times\hat{\mathbf{b}}$. Note that by using the local safety factor, defined as $q(r,\theta) = (r/R)F/\partial_r \psi$ or $\partial_r \psi = Fr/(qR)$, we can re-define this as $\mathbf{B_0} = (F/R)\left[ \hat{\bf \phi} + r/(R q) \hat{\bf \theta}\right]$ and thus $\hat{\mathbf{b}} =F/RB\left[ \hat{\phi} + r/(Rq) \hat{\bf \theta}\right]$. In this convention, the radial unit vector in torus geometry $\hat{\mathbf{r}}$ is perpendicular to $\hat{\mathbf{b}}$ and is our second coordinate. Our third theta-like coordinate, $\hat{\mathbf{\chi}}$ is defined as $\hat{\mathbf{\chi}} = \hat{\mathbf{r}} \times \hat{\mathbf{b}} = \hat{\mathbf{r}} \times F/RB\left[ \hat{\phi} + r/(Rq) \hat{\bf \theta}\right] = F/RB\left[ \hat{\bf \theta} - r/(Rq) \hat{\bf \phi}\right]$. This geometry is constructed such that $\left(\hat{\bf r}, \hat{\bf b}, \hat{\bf \chi} \right)$ and $\left(\hat{\bf r}, \hat{\bf \phi}, \hat{\bf \theta} \right)$ each form a right-handed orthogonal basis.

Then the field variables expressed in the local orthogonal coordinates,
\begin{eqnarray}
\label{eq:E_in_rchib}
    \delta\mathbf{E}&=&\delta E_r\mathbf{\hat r}+\delta E_\chi\mathbf{\hat \chi}+\delta E_\|\mathbf{\hat b}, \\
\label{eq:B_in_rchib}
    \delta\mathbf{B}&=&\delta B_r\mathbf{\hat r}+\delta B_\chi\mathbf{\hat \chi}, \\ 
\label{eq:U_in_rchib}
    \delta\mathbf{U}&=&\delta U_r\mathbf{\hat r}+\delta U_\chi\mathbf{\hat \chi},  
\end{eqnarray} 
where the compressional Alfv\'enic component ($\delta B_\parallel$) is eliminated \cite{Chen2021}. In the following derivation, field variables are also decomposed in the magnetic flux coordinates,
\begin{eqnarray}
    \delta\mathbf{E}&=&\delta{E}_r\mathbf{\hat r}+\delta \underaccent{\tilde}{E}_\theta\mathbf{\hat \theta}+\delta \underaccent{\tilde}{E}_\phi\mathbf{\hat \phi}, \\
    \delta\mathbf{B}&=&\delta{B}_r\mathbf{\hat r}+\delta \underaccent{\tilde}{B}_\theta\mathbf{\hat \theta}+\delta \underaccent{\tilde}{B}_\phi\mathbf{\hat \phi}, \\
    \delta\mathbf{U}&=&\delta{U}_r\mathbf{\hat r}+\delta \underaccent{\tilde}{U}_\theta\mathbf{\hat \theta}+\delta \underaccent{\tilde}{U}_\phi\mathbf{\hat \phi},  
\end{eqnarray} 
where the unit vector $\hat{\alpha}\equiv\nabla\alpha/|\nabla\alpha|$. The connection to Eqs. \ref{eq:E_in_rchib}--\ref{eq:U_in_rchib} is obtained readily, e.g., 
\begin{eqnarray}
\label{eq:Etheta_from_chi_par}
    \delta \underaccent{\tilde}{E}_\theta = \hat{\chi} \cdot \hat{\theta}\delta E_\chi + \hat{b} \cdot \hat{\theta}\delta E_\parallel \;\;, \\
\label{eq:Ephi_from_chi_par}
    \delta \underaccent{\tilde}{E}_\phi = \hat{\chi} \cdot \hat{\phi} \delta E_\chi +\hat{b} \cdot \hat{\phi}\delta E_\parallel \;\;.
\end{eqnarray}

\subsubsection{E and B equation in tokamak geometry}
In the geometry defined in section \ref{sec:coordinateSystem}, the momentum conservation equations of all species (Eq.\ref{eq:ChenMassflow}) is expressed as
\begin{eqnarray}
\label{eq:dUdt_tokamak0}
    &&\partial_t \left(\rho_m {U_{r,\chi}}\right) =  - (\hat{\mathbf{r}},\hat{\mathbf{\chi}})\cdot\left[ \nabla \cdot \bar\delta\bar{P} \right]_\perp +
    \frac{(\hat{\mathbf{r}},\hat{\mathbf{\chi}})}{4 \pi} \cdot \nonumber\\
    &&\left[\vec{B}_0 \cdot \nabla \delta \vec{B}_\perp  + \delta \vec{B}_\perp \cdot \nabla \vec{B}_0 +\delta \vec{B}_\perp \cdot \nabla \delta \vec{B}_\perp - \frac{1}{2} \nabla \delta B_\perp^2\right] \;\;, \label {eq:mom_conserv_all}
\end{eqnarray}
where the equilibrium has been considered ($\vec{B}_0 \cdot \nabla \vec{B}_0 - \frac{1}{2} \nabla B_0^2- \nabla \cdot \bar{\bar{P_{eq}}}=0$),  $\delta\bar{\bar{P}}=\bar{\bar{P}}-\bar{\bar{P}}_{eq}$, and the last two terms are nonlinear terms. 

The momentum conservation of all ion species are as follows,
\begin{eqnarray}
    \delta E_r &=&  \frac{1}{c} (\delta U_\chi B_0 -U_{\parallel,0}\delta B_\chi) + \frac{1}{\rho_m} \frac{m_j}{q_j} \hat{\mathbf{r}}\cdot[\nabla\cdot\delta\bar{\bar{ P}}]_\perp \label{Torus:EperpEr}\;\;,\\
    \delta E_\chi &=& -  \frac{1}{c} (\delta U_r B_0 -U_{\parallel,0}\delta B_r)
    + \frac{1}{\rho_m} \frac{m_j}{q_j} \hat{\mathbf{\chi}}\cdot[\nabla\cdot\delta \bar{\bar{P}}]_\perp\;\;, \label{Torus:EperpEchi}
\end{eqnarray}
where $U_{\parallel,0}$ is the parallel equilibrium flow and will be omitted in the following for the sake of simplicity. 

The parallel Amp\'ere's law is as follows,
\begin{eqnarray}
    &&\frac{1}{J}\frac{\partial}{\partial r}[J|\nabla r|\partial_r \delta E_\| ] 
    +\frac{1}{J}\frac{\partial}{\partial\theta} [J\frac{\hat{\mathbf{r}}\cdot\hat{\mathbf{\theta}}}{r}\partial_r\delta E_\|] \nonumber\\
    &+&\frac{1}{J}\partial_\phi[\frac{J}{R^2}(\hat{\mathbf{\chi}}\cdot\hat{\mathbf{\phi}})^2\partial_\phi\delta E_\|] 
    +\frac{1}{J}\partial_\phi[\frac{J}{Rr}(\hat{\mathbf{\chi}}\cdot\hat{\mathbf{\phi}})(\hat{\mathbf{\chi}}\cdot\hat{\mathbf{\theta}})\partial_\theta\delta E_\|] \nonumber\\
    &+&\frac{1}{J}\partial_\theta[\frac{J}{Rr}(\hat{\mathbf{\chi}}\cdot\hat{\mathbf{\phi}})^2\partial_\phi\delta E_\|] 
    +\frac{1}{J}\partial_\theta[\frac{J}{r^2}(\hat{\mathbf{\chi}}\cdot\hat{\mathbf{\phi}})(\hat{\mathbf{\chi}}\cdot\hat{\mathbf{\theta}})\partial_\theta\delta E_\|] \nonumber\\
    &-&\partial_\|\left\{
    \frac{1}{J} \partial_r[J|\nabla r| \delta E_r]
    +\frac{1}{J} \partial_\theta[\frac{J}{r} (\hat{\mathbf{r}}\cdot\hat{\mathbf{\theta}})\delta E_r] 
    \right.\nonumber\\
    &+&
    \left.
    \frac{1}{J} \partial_\theta[\frac{J}{r} (\hat{\mathbf{\chi}}\cdot\hat{\mathbf{\theta}}) \delta E_\chi]
    +\frac{1}{J} \partial_\theta[\frac{J}{r} (\hat{\mathbf{\chi}}\cdot\hat{\mathbf{\phi}})\delta E_\chi] 
    \right\}
    =\frac{4\pi}{c^2}\partial_t J_\| \;\;. \nonumber\\
\end{eqnarray}

In order to express Faraday's law, by making use of the representation $\delta E=\delta E_\alpha\nabla\alpha$ and $\nabla\times\delta E= \left( \nabla \alpha \times \nabla \beta \right)\partial_\alpha \delta{E}_\beta$, we have
\begin{eqnarray}
    -\frac{1}{c} \partial_t \delta B_r &=&
    \frac{1}{J|\nabla r|} \left( \partial_\phi \delta {E}_\theta - \partial_\theta \delta {E}_\phi \right) \;\;,\\
    -\frac{1}{c} \partial_t \delta B_\chi &=&
    \frac{\hat{\theta} \cdot \hat{\chi}}{J|\nabla\theta|} \left( \partial_r \delta {E}_\phi - \partial_\phi \delta {E}_r \right) \nonumber\\
    &+& \frac{\hat{\phi} \cdot \hat{\chi}}{J|\nabla\phi|} \left( \partial_\theta \delta {E}_r - \partial_r \delta {E}_\theta \right) \;\;, \label{eq:faraday_tokamak}
\end{eqnarray}
where $\delta E_{\theta,\phi}$ is calculated from $\delta E_{\chi,\parallel}$ as shown in Eqs. \ref{eq:Etheta_from_chi_par}--\ref{eq:Ephi_from_chi_par}. 
Equations \ref {eq:mom_conserv_all}--\ref{eq:faraday_tokamak} are general in tokamak geometry. 

Due to the decomposition of the field variables in the perpendicular and parallel directions, the equations are more complicated. In the following sections, reduced models are derived by making use of simplified geometries. For shear Alfv\'en waves, the dominant thermal ion contribution is the polarization response when energetic particles are not considered. In the following, we only include electrons as the kinetic species considering the challenge of the electromagnetic simulations with kinetic electrons and the key role of kinetic electrons in mode excitation and damping such as electron driven fish bone modes \cite{zonca2007electron}. Kinetic ions such as energetic particles can be included on the same footing. 

\subsubsection{Reduction in large aspect ratio limit with concentric circular flux surface}\label{sec:large_ratio_reduction}
In the simplified tokamak geometry with concentric circular flux surfaces, $r$ can be defined as $r=\sqrt{(R-R_0)^2+(Z-Z_0)^2}$. For an ad hoc equilibrium \cite{jolliet2009gyrokinetic}, the poloidal flux is given by $\psi(r)=\int_0^rB_0 r'/\bar{q}(r')dr'$. With a parabolic profile $\bar{q}=\bar{q}_0+\bar{q}r^2$, the analytical solution gives $\psi(r)=B_0/(2\bar{q}_2)\ln(1+\bar{q}_2 r^2/\bar{q}_0)$. In the large aspect ratio limit,
\begin{align}
    \hat{b} \cdot \hat{\phi} = \frac{F}{RB}{ \approx 1}\;\; &, \;\;
    \hat{b} \cdot \hat{\theta} = \frac{F r}{R^2 q B} {\approx \frac{r}{R q}\approx 0 } \;\;,\\
    \hat{\chi} \cdot \hat{\phi} = -\frac{F r }{R^2 q B} {\approx 0}\;\; &, \;\; 
    \hat{\chi} \cdot \hat{\theta} = \frac{F}{R B} {\approx 1}. \label{eq:phitheta2chib}
\end{align}

The Faraday's law can be simplified as follows,
\begin{align}
    -\frac{1}{c} \partial_t \delta B_r &= \frac{1}{R r} \left( \partial_\phi \delta {E}_\theta - \partial_\theta \delta {E}_\phi \right)\;\;, \label{Torus:radFaraday}\\
    -\frac{1}{c} \partial_t \delta B_\chi &= \frac{\hat{\theta} \cdot \hat{\chi}}{R} \left( \partial_r \delta {E}_\phi - \partial_\phi \delta {E}_r \right) + \frac{\hat{\phi} \cdot \hat{\chi}}{r} \left( \partial_\theta \delta {E}_r - \partial_r \delta {E}_\theta \right)\;\;. \label{Torus:chiFaraday}
\end{align}

The subscripts represent the covariant components of these variables. Considering the polarization pressure in Eq. \ref{eq:ChenPpol}, the vector calculus simplifies when considering the contravariant components. Neglecting the contribution due to ${E}_{\perp\phi0}$, we get
\begin{equation}
P_{pol} = -\frac{3}{8} \frac{q_j}{R r} \left[ \frac{\partial}{\partial r}\left( R r N_j \rho_{t,j}^2 {E}_{\perp r 0}\right) + \frac{\partial}{\partial \theta}\left(\frac{R}{r} N_j \rho_{t,j}^2 {E}_{\perp \theta 0} \right) \right]. \label{Torus:Ppol_reduced}
\end{equation}
Then we can write Eqn. \ref{eq:ChenEperp} as follows for given $P_{pol}$,
\begin{align}
    E_r &=  \frac{1}{c} U_\chi B_0 + \frac{1}{\rho_m} \frac{m_j}{q_j} \partial_r P_{pol}\;\;, \label{Torus:EperpEr}\\
    E_\chi &= -  \frac{1}{c} U_r B_0 + \frac{1}{\rho_m} \frac{m_j}{q_j} \left[  \frac{\hat{\chi} \cdot \hat{\theta}}{r}\partial_\theta P_{pol} + \frac{\hat{\chi} \cdot \hat{\phi}}{{R}}\partial_\phi P_{pol} \right].\label{Torus:EperpEchi}
\end{align}
Similarly, Eq. \ref{eq:dUdt_tokamak0} transforms into the following,
\begin{eqnarray}
    \partial_t \left(\rho_m U_r\right) &=& \frac{i}{4 \pi} \tilde{k}_\parallel B_0 \delta B_r \nonumber\\
    &-&  \frac{3}{8} q_i N_i \rho_{ti}^2 k_r \left[ k_r E_{r0} + k_\theta E_{\chi, 0}\right]\;\;,  \label{Torus:StraightMomentum_r}\\
    \partial_t \left(\rho_m U_\chi\right) &=& \frac{i}{4 \pi} \tilde{k}_\parallel B_0 \delta B_\chi  \nonumber\\
    &-&  \frac{3}{8} q_i N_i \rho_{ti}^2 k_\theta \left[ k_r E_{r0}+ k_\theta E_{\chi, 0}\right]\;\;.
\end{eqnarray}


\subsubsection{Reduction to the local model in tokamak geometry and screw pinch configuration}
Instead of solving the problem in the whole radial domain, asymptotic methods have been developed by making use of the small parameter $1/n$ and ballooning representation, where $n$ is the toroidal mode number \cite{connor1978shear,dewar1981n}. The global effects can be produced self-consistently with important properties such as mode structure symmetry breaking maintained, using the Mode Structure Decomposition approach \cite{zonca1992resonant,Lu2012,lu2017symmetry}. For physics models that are implemented in Fourier space in poloidal direction, a simple local estimate is to take the radial variation of the poloidal harmonics as an input parameter, which is very useful for estimating the excitation of Alfv\'enic instabilities in experiments \cite{lauber2009kinetic,lauber2018analytical}. 

In the local limit, the operator $\partial/\partial r$ is replaced with $ik_r$ and taken as an input. At the radial location, the mode magnitude is maximum ($k_r=0$, zero radial coupling) which is of primary interest. In screw pinch configuration, $R=R_0$ and the equilibrium variables are independent of $\theta$. The coupling between different harmonics can be ignored due to the symmetry along $\theta$.

\subsubsection{Remarks on representation in $(R,\phi,Z)$ coordinates}
While the magnetic flux coordinates $(r,\phi,\theta)$ is a physics-intuitive and natural way for the description of tokamak plasmas, as we adopted in Section \ref{sec:coordinateSystem} and in some codes  such as HMGC \cite{briguglio1995hybrid} and ORB5 \cite{lanti2019orb5}, the cylindrical coordinates $(R,\phi,Z)$ are adopted for expressing the field and/or particle equations in many code such as MEGA \cite{todo1998linear}, GT5D \cite{idomura2008conservative}, XGC\cite{chang2004numerical,chang2017fast}, JOREK \cite{huysmans2007mhd,hoelzl2021jorek} and TRIMEG \cite{Lu2019}. One advantage of the $(R,\phi,Z)$ coordinates is to treat the magnetic axis and the open field line regions as well as the X point. The GK-E\&B formulation is general and can be expressed also in $(R,\phi,Z)$ coordinates, with the similar derivation in Section \ref{subsec:rep_torus}. The derivation is straight forward but trivial and is omitted in this work. Nevertheless, the GK-E\&B fluid model is implemented in $(R,\phi,Z)$ coordinates as will be shown in Section \ref{subsec:EBfluid_circular_torus_results}.

\section{Numerical implementation}
\label{sec:numerics}
\subsection{Normalization}
 A normalization is performed, similar to that in the TRIMEG code\cite{Lu2021} using the length unit $R_N=1 \, \mathrm{m}$. The thermal velocity of a reference particle (here chosen as the electron) is adopted as the velocity unit, $v_N \equiv v_{t,e} = \sqrt{2 T_e / m_e}$. The E and B perturbed field are normalized to $E_N$ and $B_N$ where $E_N = {m_e v_{t,e}^2}/({e R_N})$, $B_N ={c \, m_e v_{t,e}}/({e R_N})$, where $e = 1.602 \times 10^{-19} \, \mathrm{C}$. The current is normalized to $J_N = e \, v_{t,N} n_0$. The factor of $c$ is included in the normalization constant $B_N$ to account for the factor of $c$ in equations \ref{eq:ChenMassflow} - \ref{eq:ChenFaraday}, although an alternative would be to include it instead in the denominator of $E_N$.

\subsection{Numerical schemes}
The GK-E\&B model is implemented in the framework of TRIMEG(-GKX). Some numerical schemes are consistent with those in the previous TRIMEG code \cite{Lu2019,Lu2021} such as the code structure of defining the equilibrium, particle and field classes, the particle-in-cell/particle-in-Fourier scheme and the coupled particle-field Runge-Kutta fourth order scheme for the local GK-E\&B model or the implicit scheme for the global GK-E\&B fluid model. However, the field equation structure of the GK-E\&B is quite different than those in the traditional $A\&\phi$ scheme where only elliptic equations are included. The time integrator for the fields for $U$ and $B$ and the field solver for $E$ are both developed newly here for the GK-E\&B. The field integrator is used with $E$ substituted from the solution of the field solver in each sub-step of the Runge-Kutta 4th order scheme. 

When implemented in $(r,\phi,\theta)$ coordinates in the local single flux surface model with kinetic electrons (Section \ref{subsec:screw_pinch_results}), the toroidal and poloidal directions are decomposed into their Fourier components $e^{i(m\theta + n \phi)}$.
When implemented in $(R,\phi,Z)$ coordinates (Section \ref{subsec:EBfluid_circular_torus_results}), the field and fluid representation is $\{E,U,B\}=\sum_{i,j,n}\{E,U,B\}_{i,j,n}e^{in\phi}N_i(R)N_j(Z)$, where $N_i(R)$, $N_j(Z)$ are basis functions in the poloidal cross section.  
For particle calculations, the particle-in-Fourier (PIF) method is used when the Fourier decomposition is adopted in the field or fluid variables in that direction, which changes the phase space integral to a discrete sum. Both the full-f and delta-f methods of PIF are implemented and used in studies of the local single flux surface model, considering that full-f may require too many particles in global simulations. The coupling between each radial surface in the full torus version is computed using the finite element method (FEM) in the poloidal cross section. Finite Larmor radius effects are neglected for simplicity of the numerical models, though they could be included through the equations presented in Reference \onlinecite{Chen2021}. For pushing variables to the next time step, the Runge-Kutta 4th order (RK4) scheme is used. All codes are written in MatLab and can be rewritten in Fortran as done before \cite{Lu2019,Lu2012} following the equations in this work.

When comparing numerical results to the analytic dispersion relation, the complex frequency of the numerical data is determined by least squares fitting to a function $f(x) = A e^{\gamma x} \sin (\omega x + \phi) + D$. For efficiency and accuracy of the fitting procedure, each variable, $f(x)$ and $x$, is normalized and eventually the obtained fitting coefficients are de-normalized once the fitting is complete.

\section{Simulation results}
\label{sec:results}
In the numerical studies, we focus on specific applications with corresponding simplifications in order to demonstrate the features of the GK-E\&B model such as its capability of capturing the Finite Larmor radius effect of thermal ions in the polarization pressure and the kinetic effects of electrons in the parallel current. The benchmark in 1D uniform plasma is performed with the comparison to the theoretical and the numerical results \cite{Chen2021} in Section \ref{subsec:slab_benchmark}. In addition, the GK-E\&B model has been implemented in two specific limits: the local single flux surface limit with kinetic electrons in Section \ref{subsec:screw_pinch_results} and the torus geometry (global) with $E_\|$ correction due to the dominant electron response in the cold electron limit in Section \ref{subsec:EBfluid_circular_torus_results}. The comprehensive implementation with all kinetic and nonlinear terms is beyond the scope of this work.

\subsection{Benchmark in uniform plasma}
\label{subsec:slab_benchmark}
Extending this problem to torus geometry requires understanding the shear Kinetic Alfv\'en Wave in uniform plasma and benchmarking our model with the previous study \cite{Chen2021}. This section validates the numerical results presented in Reference \onlinecite{Chen2021}. We consider a uniform plasma immersed in a background magnetic field $\mathbf{B} = B_0 \hat{\bf z}$ and linearize every field variable to a wave with frequency $\omega$ and wave-vector $\mathbf{k} = (k_\perp, 0, k_\|)$. We neglect the compressional KAW such that we have $\mathbf{U}_i = \delta\mathbf{U}_i$, $\mathbf{E} = \delta \mathbf{E} = (\delta E_1, 0, \delta E_\|)$, $\mathbf{B} = \mathbf{B}_0 + \delta \mathbf{B}$, $\delta \mathbf{B} = (\delta B_1, \delta B_2, 0)$. By neglecting FLR effects, letting the current be carried by electrons $J_\| \simeq J_{\|e}$, and pressure being only due to ions, it can be shown that Equations \ref{eq:ChenMassflow}-\ref{eq:ChenJparallel} reduce to the following,
\begin{align}
    & \partial_{{t}} \delta {U}_{i,2}  = i \frac{{\rho_{t,i}}}{\beta \sqrt{\tau}} \left( \frac{m_e}{m_i}\right)^{3/2} {k_\|} \, {\delta B_2}\;\;, \label{eq:SlabUi2}\\
    & {\delta E_1}  = -  \frac{1}{{\rho_{t,i}}} \sqrt{\frac{m_i}{\tau m_e}}{\delta U_{i,2}} \left( 1 -  \frac{3}{4} b_i \right)^{-1}\;\;, \label{eq:SlabE1}\\
    & {\delta E_\|} = \frac{{k_\|}}{{k_\perp}} {\delta E_1} - \frac{1}{d_e^2{k_\perp}^2} \partial_{{t}} {J_{\|,e}}\;\;, \label{eq:SlabEpar}\\
    &\partial_{{t}} {J_{\|,e}} =  {\delta E_\|} + i {k_\|} \left\langle {v_\|}^2 {F_g} \right\rangle_{e, {v}}\;\;, \label{eq:SlabJpar}\\
    &\partial_{{t}}  {\delta B_2} = -i \left( {k_\|} \, {\delta E_1} - {k_\perp} \, {\delta E_\|}\right)\;\;,\label{eq:SlabB2}
\end{align}
where $d_e^2 = \beta/\rho_{t,i}^2 \left(m_i/ m_e \right)$, $b_i = k_\perp^2 \rho_i^2/2$, $\tau = T_e/T_i$. Combining Equations \ref{eq:SlabUi2} - \ref{eq:SlabB2} leads to the following dispersion relation \cite{Chen2021},
\begin{equation}
    \frac{\omega^{2}}{\omega_{A}^{2}}=1+\frac{3}{4} b_{i}+\frac{\tau b_{i}}{1-2 \alpha_{e}^{2}+i \delta_{e}}\;\;, \label{eq:ChenDispersion}
\end{equation}
where $\omega_A =| k_\| | v_A$ and $v_A = B_0 / \sqrt{4 \pi \rho_m}$, $\alpha_e = \omega / | k_\| | v_{t,e}$, and $\delta_e = \sqrt{\pi} \alpha_e e^{-\alpha_e^2}$. 
A power series approximation of the plasma dispersion function is made here, but we use a more accurate version in the numerical implementation in Ref. \cite{zetafunctionsource}.

For numerical studies, we use equations \ref{eq:SlabUi2} - \ref{eq:SlabB2} and treat the field partial derivatives in time using the Runge-Kutta 4th (RK4) order scheme. The velocity integral $\left\langle {v_\|}^2 {F_g} \right\rangle_{e, {v}}$ is treated using either full--f Particle-in-Fourier (PIF) or delta--f PIF. In full f, $\left\langle {v_\|}^2 {F_g} \right\rangle_{e, {v}} = (1/N_{mark}) \sum_p {v}_{\|,p}^2 e^{-i{k}_\| {z}_p}$. For the delta--f approach, we first define the perturbed weight $w_1=\delta F_g/F_{g,0}$. The normalized weight equation is $\dot{w_1} = 2(1-w_1) {v}_z \dot{v}_z$. Then the parallel electron pressure term is 
calculated the weight taken into account, $\left\langle {v_\|}^2 {F_g} \right\rangle_{e, {v}} = (1/N_{mark}) \sum_p w_1 {v}_{\|,p}^2 e^{-i{k}_\| {z}_p}$.

\begin{figure}[h!]
\includegraphics[width=\linewidth]{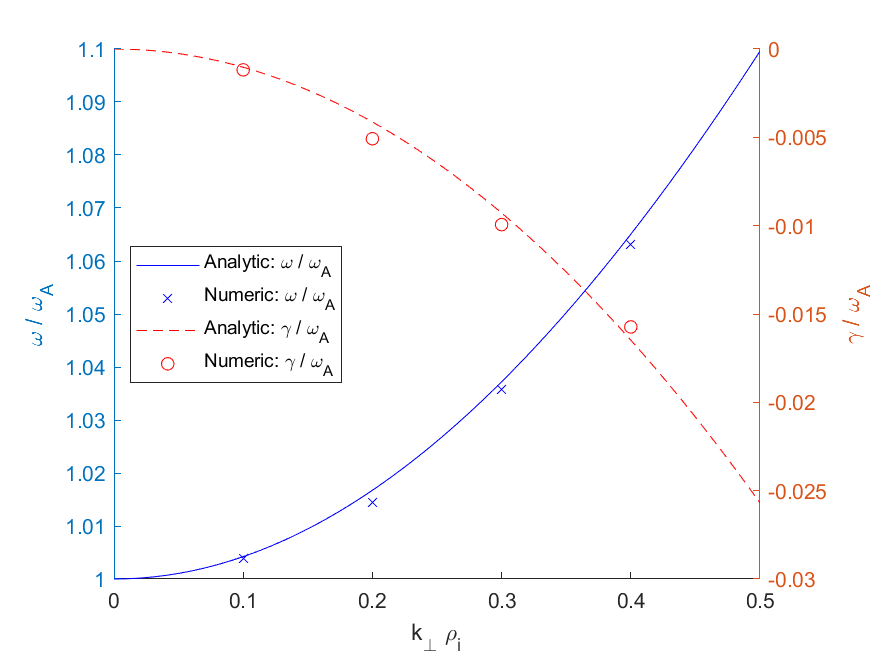}
\caption{\label{fig:1Ddispersion} Dispersion relation from the 1D problems using the full-f PIF solver. $\tau = 1$, $m_i / m_e = 1836$, $\beta = 0.01$, $k_\| / k_\perp = 10$, $N_{particles} = 10^6$, $\mathrm{dt} = T_A/100$, $\mathrm{t_{end}}=3T_A$. The simulation is performed over three Alfv\'en periods ($T_A$). The complex frequency is determined by fitting a decaying sinusoid to $\delta E_1$, $\delta U_{i,2}$, and $\delta B_2$, averaging the results together.}
\end{figure}

\begin{figure}[h!]
\includegraphics[width=\linewidth]{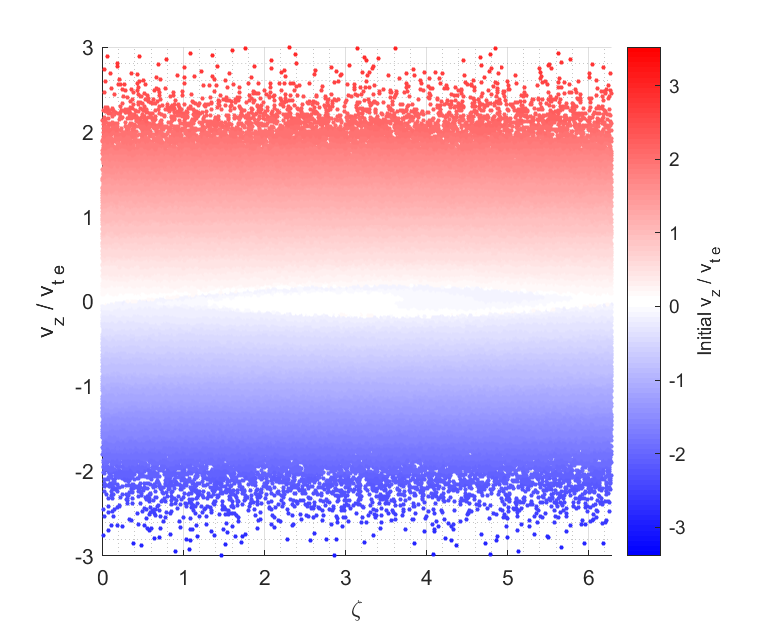}
\caption{\label{fig:1Dphase} Phase space diagram showing phase mixing. $\tau = 1$, $m_i / m_e = 1836$, $\beta = 0.01$, $k_\| / k_\perp = 10$, $N_{particles} = 10^6$, $\mathrm{dt} = 0.01 * T_A$, $k_\perp \rho_{ti} = 0.3$. The snapshot is performed at $t = 15*\omega_A$ with initial condition $\delta E_\| = 0.12$.}
\end{figure}

\begin{figure}[h!]
\includegraphics[width=\linewidth]{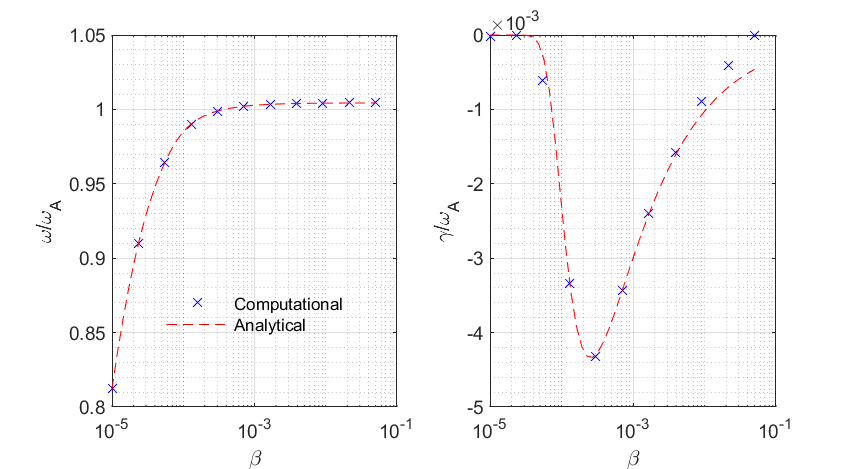}
\caption{\label{fig:1Dbetascan} Performance of the code for uniform plasma across multiple values of $\beta$ using the delta-f scheme. $\tau = 1$, $m_i / m_e = 1836$, $k_\| / k_\perp = 10$, $N_{particles} = 10^6$, $\mathrm{dt} = 0.01 * T_A$, $k_\perp \rho_{ti} = 0.1$. The left plot is frequency and the right plot is damping rate. Maximum $\beta_{max}=0.05$ for stability.}\end{figure}

The code runs without numerical instabilities and converges properly when the field and particles are semi-decoupled so the field does not see information from the particle RK4 substeps, but the particles see the field substeps. 
The normalized equations of motion for the particles reduce to the following,
\begin{equation}
    \partial_{{t}} {z} = {v_\|} \;\;, \;\;\;\;\;   \partial_{{t}} {v_\|} = -{E_\|} \;\;.
\end{equation}
Using Fourier analysis, $\partial_{{t}} {v_\|} = -2\mathrm{Re}(E_\| e^{i k_\| z})$. Periodic boundary conditions are imposed on $z$ to keep the particles within 1 wavelength. We define $\zeta = |k_\|| z$ to normalize the position to the wavelength. Results in Figures \ref{fig:1Ddispersion} and \ref{fig:1Dphase} are comparable to those in Reference \onlinecite{Chen2021}. Additionally, we show the performance of the algorithm at various values of $\beta$ in Figure \ref{fig:1Dbetascan}. Through testing, we found that the model breaks down at values of $\beta > 0.1$, $m_e/m_i=1/1863$, $k_\perp\rho_{ti}=0.1$. Numerical instabilities have not been observed for $\beta = 0.05$, which is the largest $\beta$ case shown in Figure \ref{fig:1Dbetascan}.

\subsection{Local GK-E\&B model with kinetic electrons}
\label{subsec:screw_pinch_results}

\begin{figure}[]
\includegraphics[width=\linewidth]{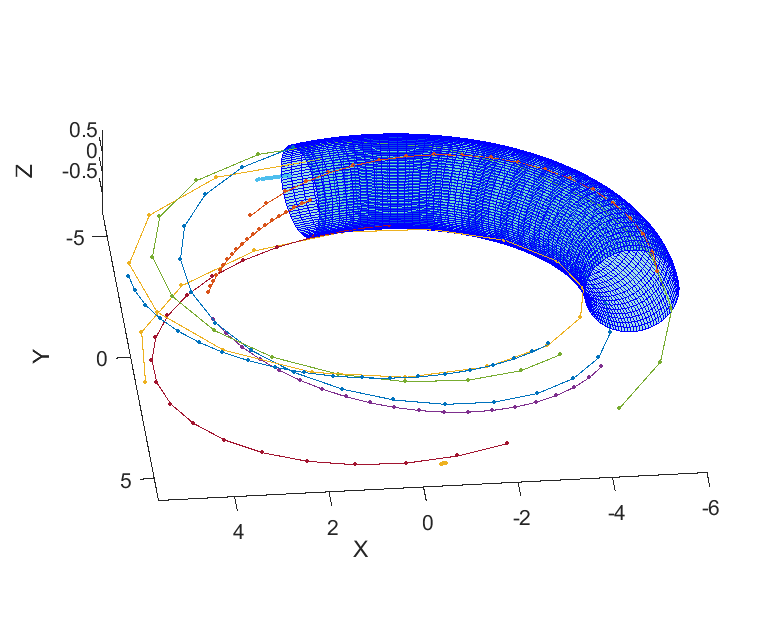}
\caption{\label{fig:2Dtrajectorytok} Particles are confined to a single flux surface and are coupled to the field.}
\end{figure}

This section considers a single shell of a torus with major radius $R$ and minor radius $a$.
The field and the particle equations are solved in the coordinate system defined in section \ref{sec:large_ratio_reduction}. 
In this simplified model, all radial coupling is ignored but kept in the derivations for generality.

In the following derivations, the results are shown in normalized quantities. Starting with Faraday's law in Eqs. \ref{Torus:radFaraday}--\ref{Torus:chiFaraday}, we substitute in Eqs. \ref{eq:Etheta_from_chi_par}--\ref{eq:Ephi_from_chi_par} to obtain the following,
\begin{eqnarray}
    \partial_t \delta B_r &=& -ik_\|\delta E_\chi +i k_\chi\delta E_\|\;\;,\\
    \partial_t \delta B_\chi &=& ik_\|  \delta E_r  - i \partial_r \delta E_\|\;\;, 
\end{eqnarray}
where $k_\| = (\mathbf{\hat{b}}\cdot\mathbf{\hat{\phi}} \,n/R + \mathbf{\hat{b}}\cdot\mathbf{\hat{\theta}} \,m/R)$ and $k_\chi = (\mathbf{\hat{b}}\cdot\mathbf{\hat{\phi}} \,m/r - \mathbf{\hat{b}}\cdot\mathbf{\hat{\theta}} \,n/R)$. For the parallel electric field, the derivation is also straightforward. The vector calculus in Eq. \ref{eq:ChenEparallel} simplifies to the following,
\begin{equation}
    \left(\partial_r^2 - k_\chi^2\right)\delta E_\| - i k_\| \left(\partial_r \delta E_r + i k_\chi \delta E_\chi \right) = \frac{1}{d_e^2}\partial_t \delta J_\|\;\;.
\end{equation}
The term $\partial_t J_\|$ takes the same form as the uniform plasma case,
\begin{equation}
    \partial_t J_\| = E_\| + i k_\| \left< v_\|^2 F_g \right>_{e,v}\;\;.\\
\end{equation}
From Eq. \ref{eq:ChenPpol}, this is simplified by considering 1 species, taking out the constants from the divergence. $\nabla \cdot \mathbf{E_\perp} = \partial_r \delta E_r + i k_\chi \delta E_\chi$, so $P_{pol}$ is the following,
 \begin{eqnarray}
    \mathbf{P}_{pol} \approx -\frac{3}{8} e n_{0i} \rho_{ti}^2 &\left[\mathbf{\hat{r}}\partial_r \delta E_r + \mathbf{\hat{\chi}}ik_\chi\delta E_\chi\right]\;\;. \label{eq:ScrewPinchPpol}
 \end{eqnarray}
 Considering Eqs. \ref{Torus:EperpEr}--\ref{Torus:EperpEchi}, using Eq. \ref{eq:ScrewPinchPpol}, we get the following, 
\begin{align}
       E_r &=  C_E U_\chi -  \frac{3}{8} \rho_{t,j}^2  \partial_r \left[ \partial_r E_r  + i k_\chi \delta E_\chi  \right]\;\;, \\
    E_\chi &= -  C_E U_r - \frac{3}{8}  \rho_{t,j}^2 i k_\chi  \left[ \partial_r E_r  + i k_\chi  \delta E_\chi \right]\;\;.
\end{align}
where $C_E = R_N/\rho_{ti}\sqrt{T_i m_i/\left(T_e m_e\right)}$. From Eq. \ref{eq:mom_conserv_all}, we neglect all the nonlinear terms and consider that along a magnetic surface, $\vec{\nabla}\delta\vec{B_0}=0$, we are only left with one term. This transforms into the following,
\begin{eqnarray}
    \partial_t \vec{U_r} &= i C_u k_\| \delta B_r +  \frac{3}{8} \frac{\overline{e_i}}{\overline{m_i}} \rho_{ti}^2  
    \partial_r \left[ \partial_r E_r 
    + i k_\chi\delta E_\chi  \right]\;\;,  \\
    \partial_t \vec{U_\chi} &= i C_u k_\| \delta B_\chi +  \frac{3}{8} \frac{\overline{e_i}}{\overline{m_i}} \rho_{ti}^2 i k_\chi\left[ \partial_r E_r 
    + i k_\chi  \delta E_\chi   \right]\;\;,
\end{eqnarray}
where $C_u = \rho_{ti}/\beta\sqrt{T_i/T_e}\left(m_e/m_i\right)^{3/2}$ and $\overline{m_i}$, $\overline{e_i}$ are the normalized ion mass and charge. Since these equations are implemented on a single flux surface, all the $\partial_r$ terms are dropped in the MatLab implementation. In this limit, these equations take the same form as the uniform plasma model. Due to this symmetry, it can be shown that the dispersion relation reduces to Eq. \ref{eq:ChenDispersion}. However, the simulation model is topologically in the torus geometry as shown in Fig. \ref{fig:2Dtrajectorytok} where the particles and fields are solved in the $(r,\phi,\theta)$ coordinates with $2\pi$ periodic boundary condition along $\phi$ and $\theta$. 

\begin{figure*}
\centering
    \subfloat[\label{fig:ExBDPower}]{\includegraphics[ width=0.49\textwidth]{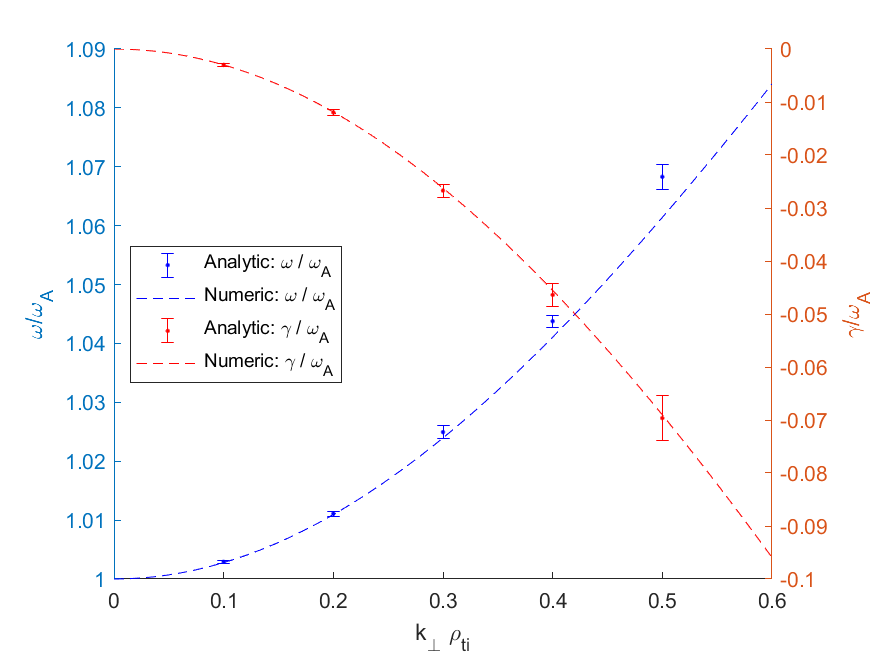}}
    \hspace{\fill}
    \subfloat[\label{fig:ExBDPower}]{\includegraphics[ width=0.49\textwidth]{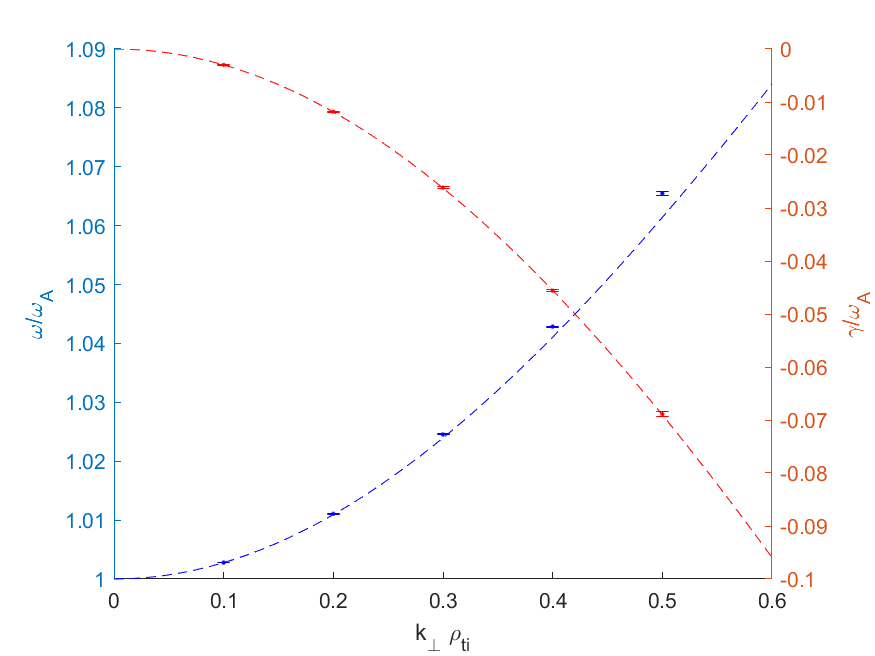}}
    \caption{\label{fig:2Dcomparef} Comparison of the local simulation dispersion relation between full--f and delta--f. $R=1$, $a=0.1$, $\beta = 0.001$, $m_i/m_e=1836$, $q=\infty$, $m,n=1$, $\tau=1$, $N_{particles}=10^6$, $\mathrm{dt}=T_A/100$, $\mathrm{t_{end}}=3T_A$. (a) full--f, (b) delta--f. Error bars represent the standard deviation over 10 trials, excluding statistical outliers.}
\end{figure*}

Our model is tested in various difficult parameter regimes to showcase its capabilities. Figure \ref{fig:2Dcomparef} highlights the differences between the full-f PIF and delta-f PIF schemes. Full-f PIF produces more run-to-run variation, whereas the delta-f approach has much less variability. However, both results are within error bars of each other. In simulations with higher values of $\beta/(m_e/m_i)$ and $k_\perp \rho_{ti}$, it is favorable to use the delta-f PIF scheme due to the full-f producing inconsistent results or needing orders of magnitude more particles. These results show excellent agreement between each other and with the theoretical predictions.

\begin{figure}[h!]
\includegraphics[width=\linewidth]{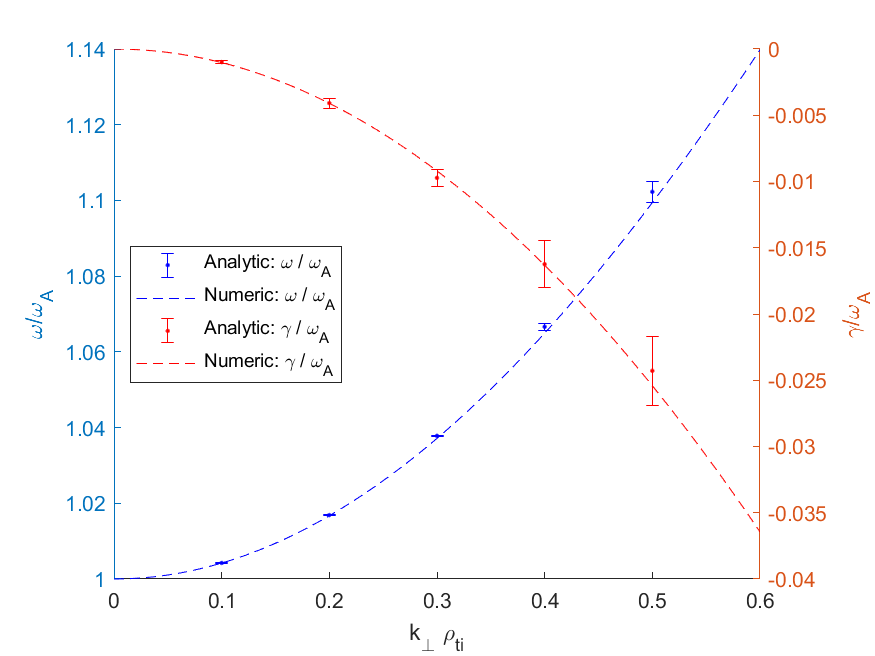}
\caption{\label{fig:2DPushItToTheLimit} The dispersion relation using the delta-f PIF solver in local screw pinch geometry. $R=1$, $a=0.1$, $q=\infty$, $\beta=0.01$, $\tau = 1$, $m_i / m_e = 1836$, $\beta = 0.01$, $m,n=1$, $N_{particles} = 10^6$, $\mathrm{dt} = T_A/100$, $\mathrm{t_{end}}=3T_A$.}
\end{figure}

\begin{figure}[h!]
\includegraphics[width=\linewidth]{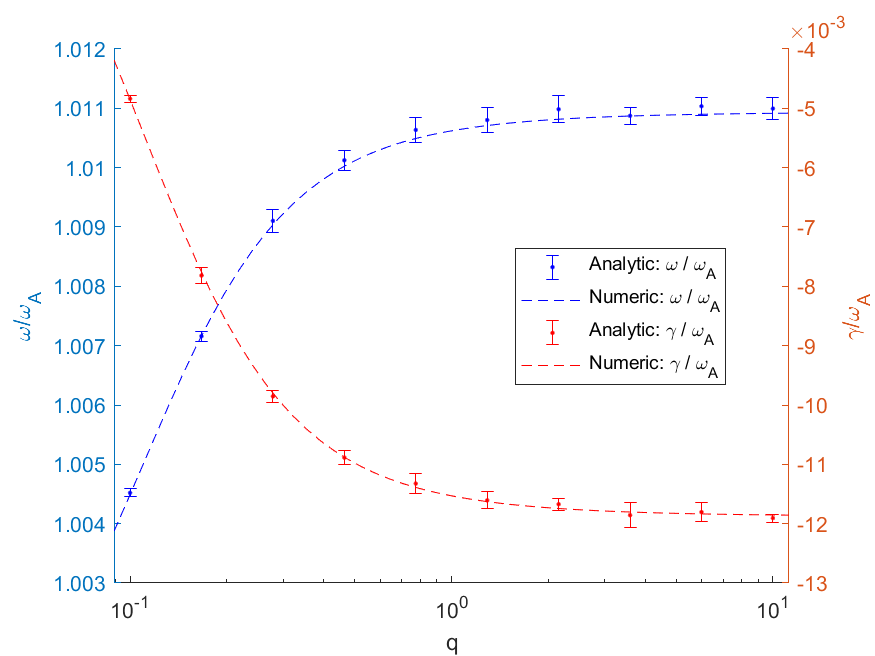}
\caption{\label{fig:2Dqscan} The real frequency and damping rate for different values of safety factor $q$ using the delta-f scheme. $T=1$, $a=0.1$, $\beta=0.001$, $\tau = 1$, $m_i / m_e = 1836$, $\beta = 0.01$, $k_\perp \rho_{ti}=0.2$ $m,n=1$, $N_{particles} = 10^5$, $\mathrm{dt} = T_A/100$, $\mathrm{t_{end}}=3T_A$}
\end{figure}

Figure \ref{fig:2DPushItToTheLimit} showcases the most stable, difficult parameter regime the delta-f PIF scheme can solve. It uses the identical conditions as shown in Figure \ref{fig:1Ddispersion}. In comparison, the local version has much more variability than the uniform plasma model. The additional variability is attributed to increasing the dimensionality to two, which reduces the particle density. Thus the code would need more particles to achieve the same accuracy of results. A stable simulation in this parameter regime is a major accomplishment. Figure \ref{fig:2Dqscan} shows that the model is correctly implemented for various values of the safety factor $q$.

In order to evaluate the performance of the GK-E\&B scheme in more realistic tokamak plasmas, another simulation is performed with parameters matched to the ITPA-TAE case. The ITPA case is defined by the parameters $R=10$, $a=1$, $q=1.75$, $m_e/m_i=1/1836$, $\beta=9\times10^{-4}$, $\tau=1$, $\rho_{ti}=0.0015205$, $m=10$, and $n=6$. The quantity $\beta m_i/(m_ek_\perp^2\rho_{ti}^2)\approx1788$ with $k_\perp\approx 20$, which is usually challenging for tradition $p_\parallel$ scheme. We use $10^6$ particles and the delta-f PIF solver. Results give a frequency $\omega/\omega_A = 1.00006063\pm 5.69*10^{-7}$ and $\gamma/\omega_A = -7.3037*10^{-5} \pm 3.333*10^{-7}$, where the error is the standard deviation over 10 trials excluding statistical outliers, compared to the theory $\omega/\omega_A = 1.00006072$ and $\gamma/\omega_A = -7.2976*10^{-5}$. The computed frequency agrees with the analytical result, only differing by $0.169$ standard deviations. The damping rate also agrees, with the theory differing by our computation by $0.183$ standard deviations. It demonstrates the capability of the GK-E\&B in simulating electromagnetic shear Alfv\'en waves in high values of $\beta m_i/(m_ek_\perp^2\rho_{ti}^2)$. 

\subsection{Global GK-E\&B fluid model with $E_\|$ in tokamak geometry}
\label{subsec:EBfluid_circular_torus_results}
The global GK-E\&B fluid model is listed in the following, as an specific application of the the original general formulae\cite{Chen2021}. Even in the most simplified form without the comprehensive kinetic part, the GK-E\&B model still takes into account important kinetic corrections, namely, the finite electron mass effect and the finite $\delta E_\|$, compared with the full MHD model. In addition, since numerous MHD or hybrid codes take the scalar/vector potential or vorticity etc as the variables instead of the E\&B field \cite{hoelzl2021jorek,briguglio1995hybrid}, it is crucial to demonstrate that in the fluid limit, the E\&B model can be implemented using specific numerical schemes, e.g., the implicit scheme in this work.  In the cold plasma limit, without considering the finite Larmor radius effects, the momentum equations Eq. \ref{eq:ChenMassflow} and Faraday's law Eq. \ref{eq:ChenFaraday}, with Eqs. \ref{eq:ChenCurrent} and \ref{eq:ChenEperp} substituted, can be written as
\begin{eqnarray}
    &\partial_t\left(\rho_m \mathbf{U}_{i \perp}\right)=\frac{1}{4\pi} (\mathbf{\nabla\times\mathbf{B}} )\times \mathbf{B} \;\;, \label{eq:fluid_mom} \\
    &\partial_{t} \mathbf{B}= \nabla  \times ( \mathbf{\mathbf{U}_{i \perp} \times \mathbf{B}-cE_\parallel})\;\;. \label{eq:fluid_faraday}    
\end{eqnarray}
Equation \ref{eq:ChenEparallel} for $E_\|$ can be simplified as
\begin{eqnarray} \label{eq:fluid_Epar}
    \left(\nabla^2_\perp-\frac{1}{d_e^2}\right) E_\parallel-{\bf b}\cdot\nabla(\nabla\cdot{\bf E_\perp})=0 \;\;,
\end{eqnarray}
where $d_e$ is the electron skin depth, and in obtaining the identity, cold electron limit $|v_{Te}k_\parallel/\omega|\ll1$ is adopted to eliminate the contribution from the electron pressure. 

\begin{figure*}[h!]
\includegraphics[width=0.65\linewidth]{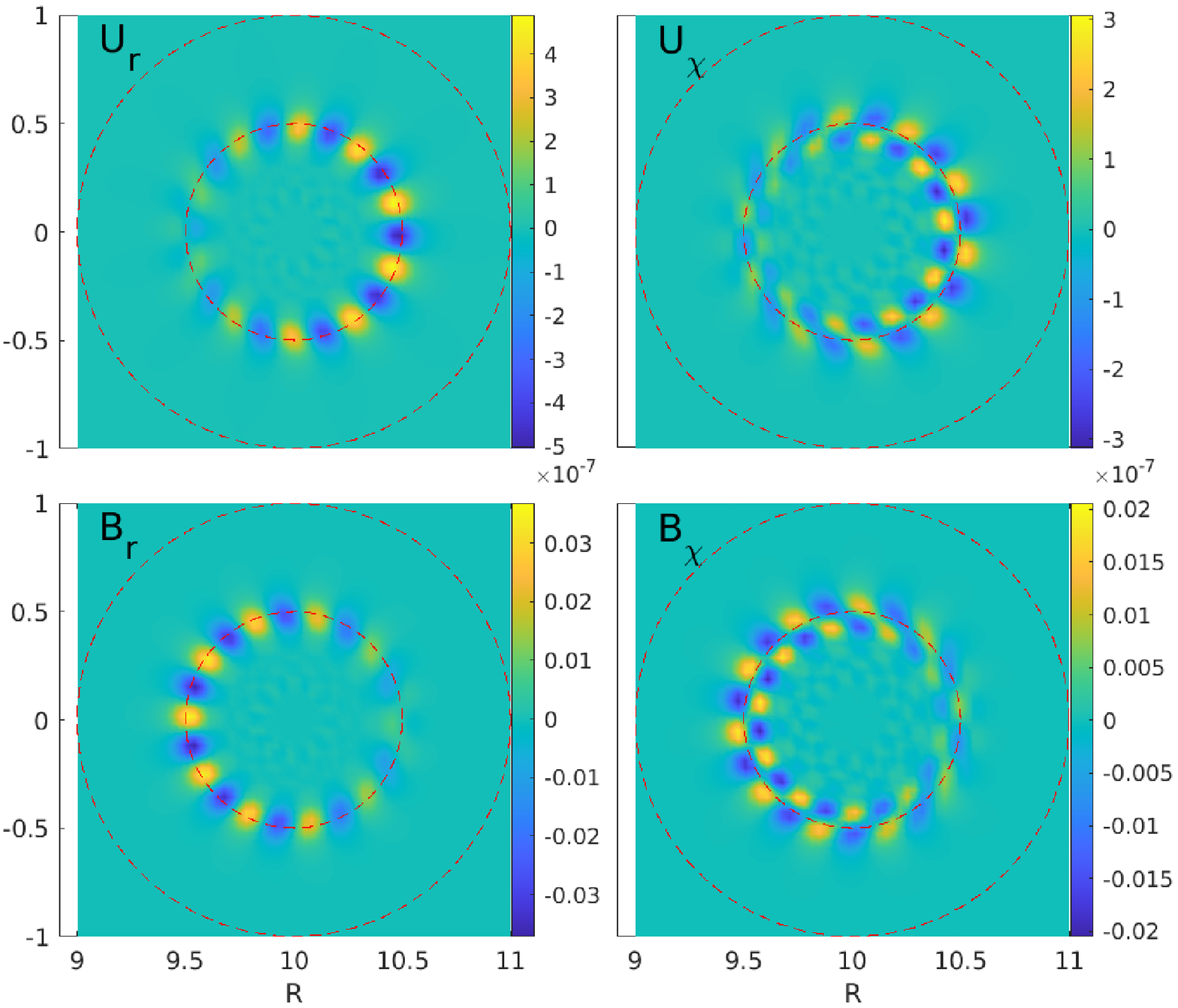}
\caption{\label{fig:UB2d}The 2D mode structures of the perpendicular component of pertubed $\delta\mathbf{\hat{U}}$ and $\delta\mathbf{B}$. Top left/right: the radial$\mathbf{\chi}$ component of the velocity perturbation; bottom left/right: the radial/$\mathbf{\hat{\chi}}$ component of the magnetic perturbation, wehre  $\mathbf{\hat\chi}=\mathbf{\hat r}\times\mathbf{\hat b}$ . Parameters are $R_0=10$, $a=1$, $q(r)=1.71+0.16r^2$ $n=6$, $\rho_{ti}=0.00152$, $m_i/m_e=3672$. All poloidal harmonics are included since no poloidal filter has been applied. }
\end{figure*}

The linearized form of Eqs. \ref{eq:fluid_mom}--\ref{eq:fluid_faraday} is solved numerically, using the cubic B-spline finite element method in the poloidal cross section but using the Fourier expansion in the toroidal direction. The formulae are derived with the same procedure as discussed in Section~\ref{sec:thoery} except that the $R,\phi,Z$ coordinate are adopted, and that the equations are adapted to an implementation friendly form, for the sake of simplicity and the convenience to treat the magnetic axis. The eigenvalues and eigen vectors are calculated in $R,\phi,Z$ coordinates. The perpendicular field and fluid perturbation $\delta\mathbf{U}$ and $\delta\mathbf{B}$ are as shown in Fig. \ref{fig:UB2d}. For this ITPA-TAE case, in the previous benchmark work \cite{konies2018benchmark}, the scalar potential $\delta\phi$ has been compared among various codes, demonstrating the ballooning structure of the 2D mode structures in $r,\theta$ plane, namely, $\delta\phi_{m=10,11}(r)\sim \exp\{-(r-r_m)^2/W_m^2\}$, where $W_m$ indicates the width in the radial direction. The corresponding $\delta U_r$ and $\delta U_\chi$ are also expected to be ballooned at the low field side, with the radial node number being $0$ and $1$ respectively, since $\delta\mathbf{U}=\mathbf{B}_0\times\delta\mathbf{E}$. The 2D structure of $\delta U_r$ and $\delta U_\chi$ shows this feature and is consistent with the one obtained in other codes \cite{konies2018benchmark}. The perturbed magnetic field shows anti-ballooning structures, which is consistent with that of $\delta A_\parallel$ in gyrokinetic simulations \cite{biancalani2016linear}. Since most codes have not taken $\mathbf{E}$ and $\mathbf{B}$ as variables, closer benchmark will require more dedicated efforts and more quantitative comparison will be addressed in the future. 

\begin{figure*}[h!]
\includegraphics[width=0.98\linewidth]{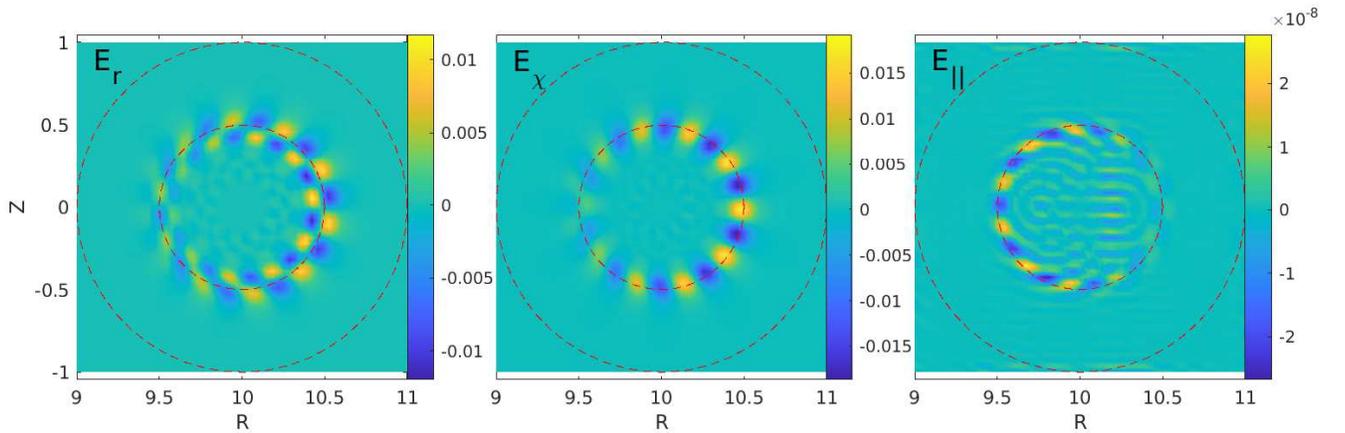}
\caption{ \label{fig:E2d}The 2D mode structures of perturbed $\mathbf{E}$. The component in the radial (left), along $\mathbf{\hat\chi}=\mathbf{\hat r}\times\mathbf{\hat b}$ (middle) and parallel (right) direction. The parameters are the same as Fig. \ref{fig:UB2d} }
\end{figure*}

\begin{figure}[h!]
\includegraphics[width=0.88\linewidth]{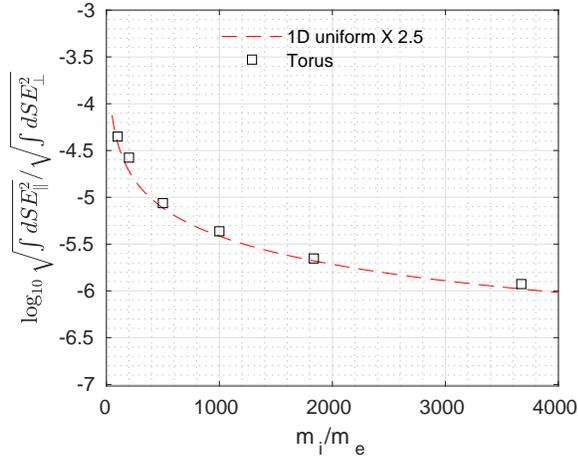}
\caption{\label{fig:EparEperVSMe}The root mean square (RMS) of the parallel electric field divided by the RMS of the perpendicular electric field versus the ion-electron mass ratio. The parameters are the same as Fig. \ref{fig:UB2d} except $m_i/m_e$ which is the scan parameter. }
\end{figure}

An important feature of the GK-E\&B model is the treatment of the finite electron mass effect in the $E_\parallel$ equation. In the cold plasma limit, the $\partial_t J_{e,\parallel}$ term is substituted with the dominant contribution $E_\|/d_e^2$ in Eq. \ref{eq:fluid_Epar}. The eigen solution of $\mathbf{E}$ is shown in the left and middle frames of Fig. \ref{fig:E2d}, where the ballooning structures of $\delta E_r$ and $\delta E_\chi$ are consistent with those from other code \cite{konies2018benchmark}. In addition, $\delta E_\|$ is also produced although its magnitude is $10^{-6}$ of $\delta E_{r,\chi}$, as shown in the right frame of Fig. \ref{fig:E2d}. In the zero electron mass limit, $\delta E_\|$ is zero. As finite electron mass is taken into account, finite $\delta E_\|$ is generated. As shown in Fig. \ref{fig:EparEperVSMe}, as the electron mass increases, the ratio between $\delta E_\|$ and $\delta E_\perp$ increases, where the average magnitude of $\delta E_{\|,\perp}$ is estimated using the root mean square of the field variable in the whole volume. The red dashed line is the result using the uniform model in the cold plasma limit, with a factor $2.5$ multiplied, in order to match the results in the torus geometry. The parameters for the uniform model are $k_\perp\approx m/2=21$, $k_\parallel=1/(2q_cR_0)$, $q_c=1.75$, $R_0=10$, $\rho_i=0.00152$, $\beta=9\times 10^{-4}$. The torus model and the uniform model are different, such as the ballooning/anti-ballooning structures of the 2D model structure and varying effective radial wave vector $k_r$ at each location and the varying radial amplitude of the mode structure, which can lead to the difference between $\sqrt{\int dS\delta E_\|^2}/\sqrt{\int dS\delta E_\perp^2}$ in torus and $\delta E_\|/\delta E_\perp$ in the uniform plasmas. Nevertheless, the key physics ingredient of the finite mass effect is included in both cases and the uniform model provides a good estimate. For realistic electron mass, $|\delta E_\||$ is very small ($|\delta E_\||/|\delta E_\perp|\sim 10^{-6}$) but is the key ingredient for simulating the electron Landau damping, as is analyzed in uniform plasma numerically and theoretically \cite{Chen2021}.

\begin{figure*}[h!]
\includegraphics[width=0.42\linewidth]{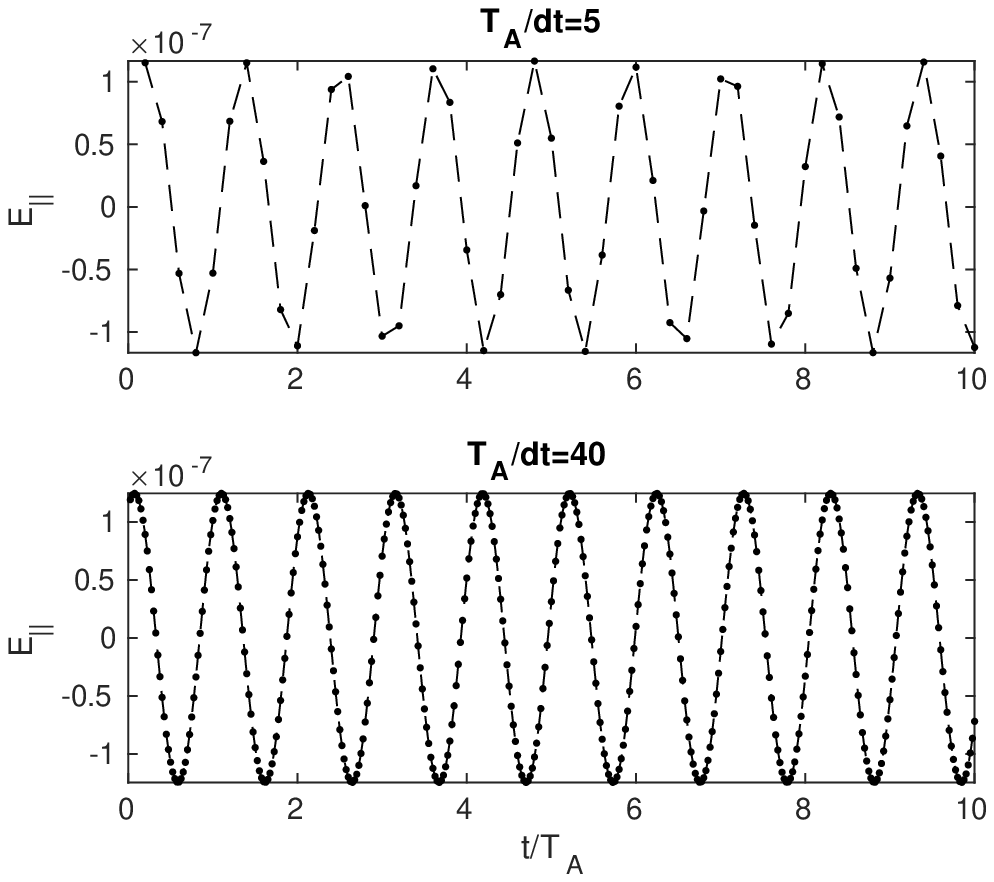}
\includegraphics[width=0.44\linewidth]{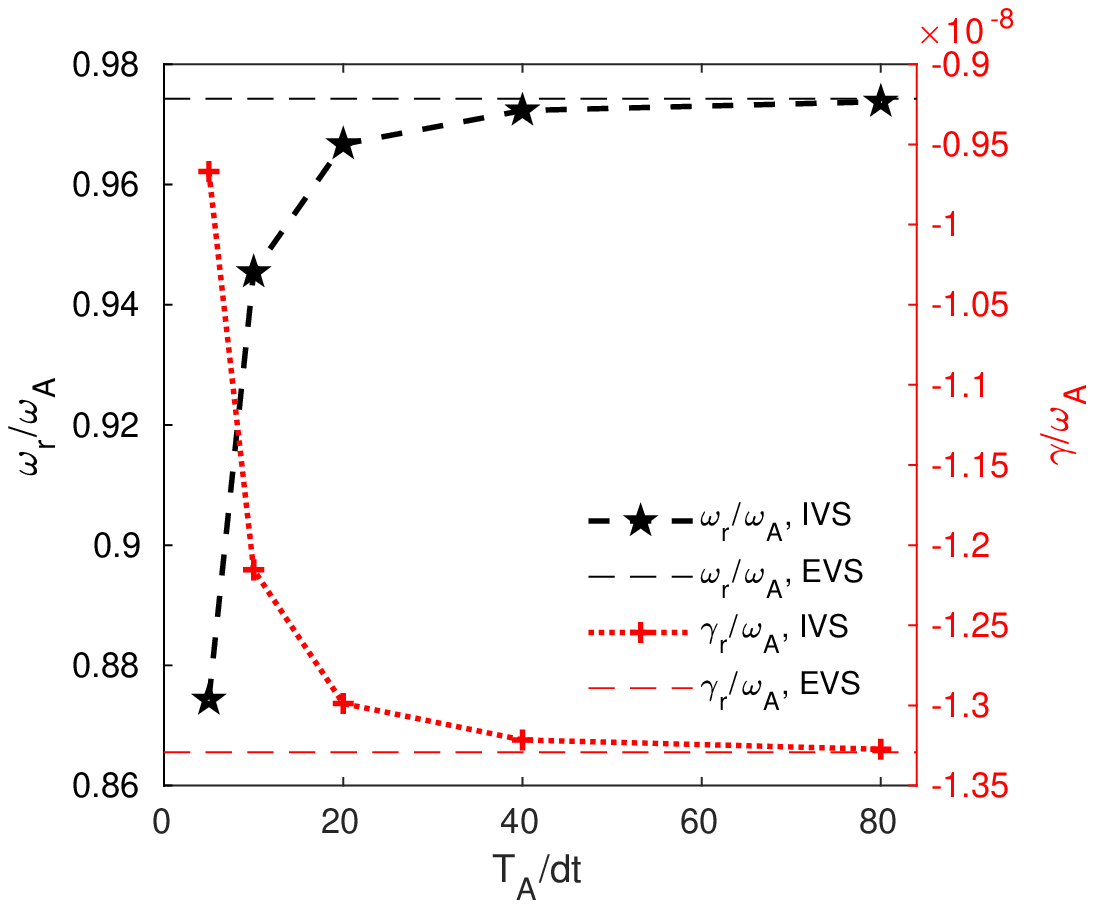}
\caption{\label{fig:EparVSt}Left: the time evolution of $\delta E_\|$ at the probe location where the magnitude is maximum. Right: the fitted real frequency and damping rate from the initial value solver (markers) versus step number per wave period. The dashed lines are the real frequency (black) and the damping rate (red) from the eigenvalue solver. The parameters are the same as Fig. \ref{fig:UB2d} except $m_i/m_e=1836$.}
\end{figure*}

An implicit initial value solver has been implemented in addition to the eigenvalue solver. As shown in the left frame of Fig. \ref{fig:EparVSt}, the evolution of $\delta E_\|$ is simulated with only 5 steps per wave period chosen. The frequency and the imaginary part of the mode is fitted and compared to the eigenvalue solution $\omega/\omega_{TAE}=0.974242461257867 - 1.3292321\times 10^{-8} i$, where $\omega_{TAE}=0.5/(q_cR_0)$, $q_c=1.75$, as shown in the right frame of Fig. \ref{fig:EparVSt}. The imaginary part of $\omega$ is from numerical error and is much smaller than any physics damping mechanism of interest. The implicit initial value solver gives converged results in terms of real frequency as $T_{TAE}/\Delta t\ge 40$. $\delta E_\|$ evolves properly in the initial value solver even its magnitude is much smaller than $\delta E_\perp$. While in this work we focus on the studies of the basic features of the GK-E\&B scheme, the coupling to the kinetic particles relies on more dedicated numerical techniques such as parallelization and advanced field solvers, and will be implemented in the future, for more comprehensive studies of tokamak plasmas.

\section{Conclusion and outlook}
\label{sec:conclusion}

In this work, the equations of the E and B gyrokinetic model are derived in screw pinch configuration and tokamak geometry. Local orthogonal coordinates are adopted for representing the field variables while tokamak flux coordinates are adopted for expressing the operators. 
The reduction of the general equations in the screw pinch and large aspect ratio tokamak geometry with concentric circular magnetic flux surfaces are demonstrated.
Numerical benchmarks in a uniform plasma show good agreement with previous work \cite{Chen2021}.
The model is then implemented in the screw pinch configuration and tokamak geometry.
Numerical results of the frequency and the damping rate of the kinetic Alfv\'en wave agree with the theoretical results of the dispersion relation derived in screw pinch configuration.
With this model implemented in tokamak geometry for concentric circular geometry, the TAE is simulated with GK-E\&B pure fluid model, with the finite parallel electric field simulated in the cold electron limit. 
The simulation results using the ITPA-TAE parameters show reasonable agreement with the previous work in terms of mode structures and eigenvalue w/o energetic particles. 
With the capability of treating the electromagnetic effect and the parallel dynamics of electrons, the E and B gyrokinetic model provides a good choice for simulating burning plasmas featuring high values of $\beta m_i/(m_e k_\perp^2\rho_{ti}^2)$. Moreover, the GK-E\&B model can be implemented for even higher frequency $\omega\ll\Omega_e$ (electron cyclotron frequency) in tokamak geometry following the original theoretical framework \cite{Chen2019} and the treatment in this work, which enables the studies in a much broader frequency regime (such as lower hybrid frequency $\sim GHz$) that can not be handled by most of the existing gyrokinetic codes.


While in this work, we are focusing on the derivation of the GK-E\&B model in tokamak geometry and the numerical analyses of its basic properties, more comprehensive studies are needed by more dedicated development of the GK-E\&B code, with more ingredients such as the realistic geometry, magnetic equilibrium and experimental profiles. The coupling between the fluid part and the kinetic part is also needed in the global torus version, in order to take into account the kinetic closure in the fluid equations. The combination of the GK-E\&B with the unstructured meshes and finite element method \cite{chang2017fast,Lu2019} can enable the studies including magnetic axis and open field line regime and can provide powerful tools for studies of edge physics, energetic particle physics and burning plasma dynamics, which merits more efforts in the future. 




\section*{Acknowledgments}
Simulations in this work were partly performed on Max Planck Computing \& Data Facility (MPCDF). Discussions with and inputs from R. Hatzky on numerical methods are appreciated by ZL. This work is supported by the Fulbright program. 
This work has been carried out within the framework of the EUROfusion Consortium and has received funding from the Euratom research and training programme 2014-2018 and 2019-2020 under grant agreement No 633053. The views and opinions expressed herein do not necessarily reflect those of the European Commission. 
\\
	
\nocite{*}
\providecommand{\noopsort}[1]{}\providecommand{\singleletter}[1]{#1}%

\end{document}